\pdfoutput=1

\documentclass[11pt]{article}

\usepackage[]{naacl2021}

\usepackage{times}
\usepackage{latexsym}

\usepackage[T1]{fontenc}

\usepackage[utf8]{inputenc}
\usepackage{graphicx}

\usepackage{microtype}

\title{IITD-DBAI: Multi-Stage Retrieval with Pseudo-Relevance Feedback and Query Reformulation}

\author{Shivani Choudhary\\ shivani@sire.iitd.ac.in \\ Indian Institute of Technology, Delhi}

\begin{document}
\maketitle


\section{Introduction}
Conversational systems have acquired the center stage in NLP research. Compared to the conventional information retrieval task where we have to extract the passage or document from a vast collection of documents, the Conversational system requires extracting related information to respond to a series of questions. The turns in the conversation may follow the previous question. Complexity in this task arises due to the way we form the queries, which often have a reference to previous information using pronouns, co-reference. The presence of pronouns and unresolved co-references induces ambiguity in the query. Resolving the contextual dependency is one of the most challenging tasks in the Conversational system. 

The Conversational assistance track (CAsT) has started in 2019. The long-term vision of the CAsT is ``to support natural conversations between a person and a search engine to satisfy information needs and support complex information tasks''. The task in both years of CAsT remained the same, to retrieve relevant passages depending upon the context evolution from the subsequent queries. 
raw\_utterance
CAsT-2019 had two tacks viz: automatic and manual track. Under Automatic track, response extraction was based on the raw utterances while the manual track has queries rewritten by humans. Co-reference and pronouns were resolved in the manually rewritten queries based on the historical context. Manually rewritten queries contain all of the information required to represent the single turn of the underlying information need. In CAsT-2019, additional information like description and title for the session was also provided \citep{Dalton2020}. Three corpora, namely MSMARCO, TREC CAR (Wikipedia) paragraph Corpus V2.0\footnote{http://trec-car.cs.unh.edu/datareleases/} were used. Initially, it had also considered WaPo, but it was later dropped. CAsT-2020 had some changes, title and description was \textbf{removed from the query} info, and document id as a canonical response to the query is added in the task. CAsT-2020 had three tracks - automatic, automatic-canonical, and manual. Under automatic, only raw utterance is to be used, while automatic-canonical can use the provided canonical response. The manual track was similar to last year based on the manually rewritten queries \citep{Dalton2021}. The dataset was MSMARCO and CAR \citep{Dietz2018TRECCA}. CAsT-2021 has three tracks similar to CAsT-2020, but with a modification that we need to extract a specific passage out of the extracted document. The idea of specific passage to be extracted from the document centered around the idea that responses should be crisp and could be used by automatic voice assistants like Alexa, Google, etc. CAsT-2021 used MSMARCO (2019/20 dump), WAPO-2020 and KILT \citep{petroni-etal-2021-kilt}.  

\begin{table}[]
    \centering
    \begin{tabular}{|l|p{0.8\linewidth}|}
         \hline
         Turn & Queries  \\
         \hline
         1 & I just had a breast biopsy for cancer. What are the most common types? \\
         \hline
         2 & Once it breaks out, how likely is it to spread?\\
         \hline
        3 &	How deadly is it?\\
        \hline
        4 &	What? No, I want to know about the deadliness of lobular carcinoma in situ. \\
        \hline
        5 & Wow, that's better than I thought.  What are common treatments? \\
        \hline
        6 & ... \\
        \hline
    \end{tabular}
    \caption{A sample query from CAsT-2021 Automatic Evaluation Topics }
    \label{tab:my_label}
\end{table}

\section{Problem Description}
In CAsT track, A conversation session $S$ has a series of utterances $\{u_1, u_2, u_3, u_4\dots\}$ called as turns. Our task is to predict a set of top-K passages from the collection for each turn. 

\textbf{Tracks:} There are three tracks of submission for CAsT-2021 - Automatic, Automatic-canonical, and Manual. Manual track has rewritten queries as input.

\textbf{Dataset}: CAsT-2021 uses two dataset corpus for the task - MSMARCO, KILT \citep{petroni-etal-2021-kilt}  and WAPO. 

We have attempted the Automatic track. Each turn (Queries) in this session is raw in nature. The onus to resolve references and pronouns lies on the model itself. 

\section{Motivation and Method}
We have gone through the CAsT-2020 submissions, and most of the models were multistage models. In the initial stage, a set of documents were retrieved from the collection, followed by re-ranking using neural model. Most of the models used query rewriting using generative model based on Transformer architecture \citep{DBLP:journals/corr/VaswaniSPUJGKP17}. For query rewriting, T5, BART, GPT-2 was used, and the re-ranker engine was based on BERT, ALBERT, and T5.  

Our retrieval framework has three components: Query rewrite, document retrieval with pseudo-relevance feedback, and neural engine-based re-ranker. The Query rewrite framework is based on chatty-goose \footnote{https://github.com/castorini/chatty-goose} and re-ranker is based on pygaggle \footnote{https://github.com/castorini/pygaggle}. Core retrieval engine is based on the HQE \citep{Yang2019QueryAA} and PQE  \citep{Al-Thani2021}
\begin{figure}
    \centering
    \includegraphics[width=\linewidth] {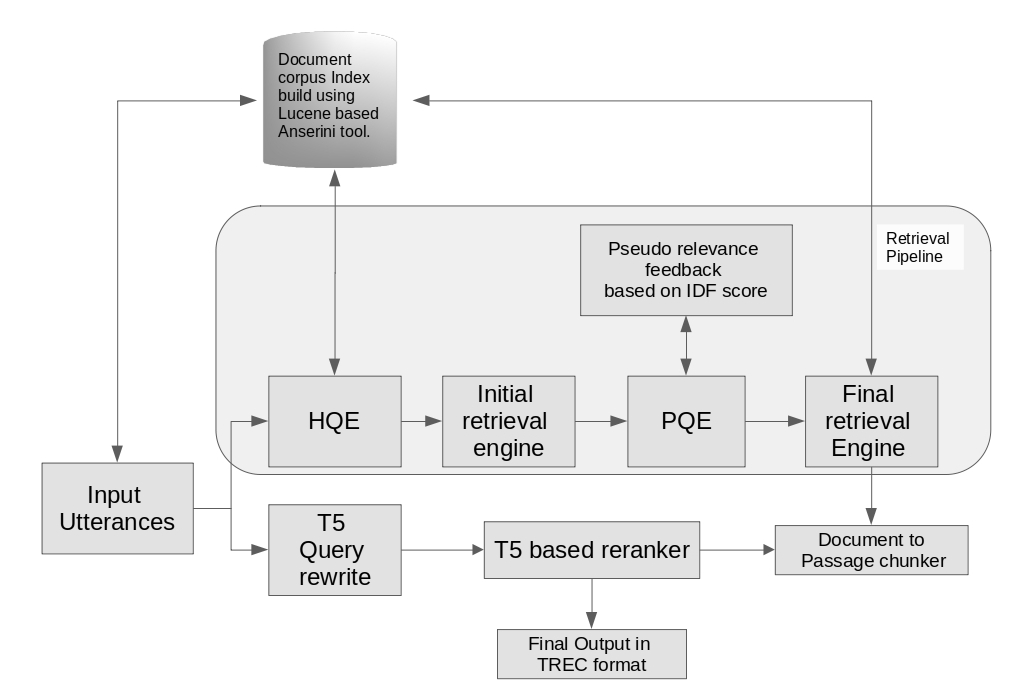}
    \caption{Multi-stage retrieval pipeline}
    \label{fig:my_label}
\end{figure}

\subsection{Historical Query expansion (\textbf{HQE})}
This step is based on the submission from CAsT-2019 \citep{Yang2019QueryAA}. It is a three-stage algorithm. In the first step, it extracts the keyword for the session and query level followed by measurement of ambiguity in query in second step, and in the last stage, query expansion with the session keyword and query level keyword is done. 

A keyword is deemed important if it strongly relates with the documents in the index. HQE uses BM25 score between a keyword and its highest-ranked document in the index as its measure of importance. HQE computes this measure for each token in every utterance when presented with a topic. The most informative ones are then selected for query expansion.
For a given topic, keywords can be locally important (i.e., it strongly relates to the topic being discussed in the current utterance) or globally important (i.e., it relates to the overall conversation theme). HQE thus creates two sets of expansion keywords, session and query, during the extraction phase.
Whether the keyword is considered a session keyword or a query keyword is decided by two separate cutoffs, $Q_s$, and $Q_t$. If the importance score of a keyword is greater than either of the cutoff scores, it is added to corresponding expansion sets. Note that session keywords are always used for expansion, while query keywords are only used when current utterance is identified as ambiguous. An utterance is deemed ambiguous when BM25 score between it and its highest ranked document in index is low (i.e. by itself utterance is not important). If the ambiguity score is less than a certain threshold $\theta$, then the query expansion will take place with query keywords.

The cutoff values for $Q_s$, $Q_t$, and $\theta$ determine the performance of HQE. We follow a greedy approach to find the optimal values of these cutoffs. Note that a query is always expanded with current session keywords and occasionally with preceding query keywords. Thus, we start by tuning $Q_s$ session cutoff. We set $Q_t$ and $\theta$ to arbitrarily high values allowing query expansion only via session keywords. Then we do a line search for $Q_s$ over the training set. Second, we set $Q_t$ to some fixed value ($> Q_s$) and in similar fashion tune $\theta$. Finally, we tune $Q_s$. 

\begin{table*}[]
    \centering
    \begin{tabular}{p{0.4\linewidth}|c|c|c|c|c|c}
        \hline
        Run Name & $K1$ & $b$ & NDCG@3 & NDCG@5 & P@5 & AP@500\\
        \hline
        IITD-RAW\_U\_T5\_1 & $0.9$ & $0.4$ & 0.3712 & 0.3631 & 0.5025 & 0.1759\\
        \hline
        IITD-RAW\_U\_T5\_2 & $1.2$ & $0.75$ & 0.3801 & 0.3731 & 0.5203  & 0.1874 \\
         \hline
    \end{tabular}
    \caption{Evaluation score of Submitted manual runs}
    \label{tab:my_label_2}
\end{table*}

\subsection{Passage Query Expansion (PQE)}
Pseudo-relevance feedback enriches the query by incorporating features from top-k relevant documents. PQE uses a pseudo-relevance feedback mechanism as follows:
\begin{itemize}

\item The expanded query is from HQE is used to fetch an initial set of top-k documents. These documents combined together form the corpus of responses for the query.
\item Individual tokens in the corpus are scored using TF-IDF. An IDF vector is pre-computed on complete MSMARCO documents.
\item The topmost unique tokens are then considered for query expansion.
\end{itemize}
Note that PQE can be computationally prohibitive since it involves complete document retrieval. In order to avoid this, \cite{Al-Thani2021} proposes to use a simple rule to decide whether a query is to be expanded using PQE or not. \cite{Al-Thani2021} only perform PQE expansion when the query has at least one pronoun.

This process has two HyperParameters, top-k documents, and top-k tokens from the document based on the TF-IDF score. The TF-IDF score helps the system to select the important terms. While selecting the term, we have also placed a criterion that it should have a DF between 0.001 and 0.2. In the PQE step, we noticed that some digits also appeared in top-k terms. Later, we filter out those instances.


\subsection{T5 query rewriter and reranker}
In automatic track, query are presented in the bare format. In order to preserve the context of the query, we have used query rewriter. It uses T5 based model \footnote{https://huggingface.co/castorini/t5-base-canard}. The model is trained on CANARD dataset \citep{Elgohary:Peskov:Boyd-Graber-2019} that contains a set of rewritten queries based upon the history. To produce a $n^{th}$ rewitten utterance, we have supplied history of turns $\{u_1, u_2, u_3, u_4\dots, u_{n-1}\}$ concatenated with $u_n$. Reformulated queries were used for the final stage ranking of passages. 

On the other hand T5 based ranking model \citep{DBLP:journals/corr/NguyenRSGTMD16} is trained on the MSMARCO, Robust04, Core17 and Core18. This model takes query $Q$ and a list of passages $\{p_1, p_2, p_3, .. p_n\}$. It returns submitted passages according to the descending order of their relevance.  

\subsection{Document Indexing}
All three datasets were pre-processed and converted into jsonl format. We have use Pyserini\footnote{https://github.com/castorini/pyserini} to generate an index for faster retrieval of documents. Index was generated with an option to keep a copy of raw documents. We choose this option because the raw content was required at HQE and PQE stage.

\section{Evaluation Matrix}
CAsT-2021 used following matrix to present the results of the participants NDCG@3, NDCG@5, NDCG@500 and AP@500.

\section{Results and Discussion}

Index for the cleaned document was generated using Pyserini\footnote{https://github.com/castorini/pyserini} with the default setting. Default setting did not restrict us to keeping $K1$ and $b$ fixed. It can be changed during the retrieval of the documents. First, we performed our extraction and tuning on the 2019 training set. The best performance for HQE was obtained with $Q_s$ = 4, $Q_t$ = 4, $\theta$ = 10 and PQE with top-k = 5, top-k token = 3.  

We have participated in automatic track and submitted two runs in the CAsT-2021. Two different set of parameters for BM25, $K1 = 0.9$, $b = 0.4$ and $K1 = 1.2$, $b = 0.75$ was selected for the retrieval of the document. The MAP is very low in both the results, and the main issue lies with the recall. Our retrieval engine extracted the top 100 documents from the corpus. Extracted documents were later chunked and re-ranked. In this process, lower retrieval numbers left the ranker with fewer relevant chunks, resulting in a lower than expected performance. 

IITD-RAW\_U\_T5\_2 has produced a mean NDCG@3 performance better than the median model. Model's performance on evaluation query has a noticeable variation in scores compared with median scores. Results for queries like -- 115 and 119 were better than the median; however, results on the evaluation queries like -- 111 and 117 were worse than the median benchmark for NDCG@3. Analysis of query expansion term for the first stage retrieval suggests that query expansion term has carried the intent of the conversation to higher depths. While the poor performing queries have expansion terms with less relevant keywords, generic keywords. Query expansion terms for query 119 have terms like \textit{swelling, shaking, infection, ear} that carried the conversation context to higher depth led to better performance. 

We have also analyzed the hqe and pqe keywords for query expansion. We can take evaluation number 106; the first few queries are listed in Table-\ref{tab:my_label}. For turn 3, query expansion terms were \textit{biopsy, breast, deadly, cancer, comment and cell}. This turn has only one term, ``deadly'', that can explain the intent of the query. But, it is too generic in nature which has led to poor recall. On the flip side, turn five is expanded with terms \textit{carcinoma, wow, deadly, situ, lobular, deadliness, biopsy, breast} which has specific terms related to cancer like ``lobular, situ, carcinoma'' has a good recall. Here the word ``specific'' means relevance to the topic. 

Our submission to CAsT-2021 aimed to preserve the key terms and the context in all subsequent turns and use classical Information retrieval methods. It was aimed to pull as relevant documents as possible from the corpus. It appears that it can retain some of the keywords in subsequent turns. But, it fails when the key term itself is generic. The performance of this model can be improved further by including the context word vectors in the HQE and PQE stages. Context vector-based selection may help to keep top-k terms that are relevant to the conversation theme.

\bibliography{anthology,custom}

\begin{thebibliography}{9}
\expandafter\ifx\csname natexlab\endcsname\relax\def\natexlab#1{#1}\fi

\bibitem[{Al-Thani et~al.()Al-Thani, Jansen, and Elsayed}]{Al-Thani2021}
Haya Al-Thani, Bernard~J Jansen, and Tamer Elsayed.
\newblock \href {https://doi.org/10.1145/nnnnnnn.nnnnnnn} {{HBKU at TREC 2020:
  Conversational Multi-Stage Retrieval with Pseudo-Relevance Feedback; HBKU at
  TREC 2020: Conversational Multi-Stage Retrieval with Pseudo-Relevance
  Feedback}}.

\bibitem[{Dalton et~al.(2020)Dalton, Xiong, and Callan}]{Dalton2020}
Jeffrey Dalton, Chenyan Xiong, and Jamie Callan. 2020.
\newblock \href {http://arxiv.org/abs/2003.13624v1} {{CAsT 2019: The
  Conversational Assistance Track Overview}}.
\newblock Technical report.

\bibitem[{Dalton et~al.(2021)Dalton, Xiong, and Callan}]{Dalton2021}
Jeffrey Dalton, Chenyan Xiong, and Jamie Callan. 2021.
\newblock \href {https://huggingface.co/} {{CAsT 2020: The Conversational
  Assistance Track Overview}}.
\newblock Technical report.

\bibitem[{Dietz et~al.(2018)Dietz, Verma, Radlinski, and
  Craswell}]{Dietz2018TRECCA}
Laura Dietz, Manisha Verma, Filip Radlinski, and Nick Craswell. 2018.
\newblock Trec complex answer retrieval overview.
\newblock In \emph{TREC}.

\bibitem[{Elgohary et~al.(2019)Elgohary, Peskov, and
  Boyd-Graber}]{Elgohary:Peskov:Boyd-Graber-2019}
Ahmed Elgohary, Denis Peskov, and Jordan Boyd-Graber. 2019.
\newblock Can you unpack that? learning to rewrite questions-in-context.
\newblock In \emph{Empirical Methods in Natural Language Processing}.

\bibitem[{Nguyen et~al.(2016)Nguyen, Rosenberg, Song, Gao, Tiwary, Majumder,
  and Deng}]{DBLP:journals/corr/NguyenRSGTMD16}
Tri Nguyen, Mir Rosenberg, Xia Song, Jianfeng Gao, Saurabh Tiwary, Rangan
  Majumder, and Li~Deng. 2016.
\newblock \href {http://arxiv.org/abs/1611.09268} {{MS} {MARCO:} {A} human
  generated machine reading comprehension dataset}.
\newblock \emph{CoRR}, abs/1611.09268.

\bibitem[{Petroni et~al.(2021)Petroni, Piktus, Fan, Lewis, Yazdani, De~Cao,
  Thorne, Jernite, Karpukhin, Maillard, Plachouras, Rockt{\"a}schel, and
  Riedel}]{petroni-etal-2021-kilt}
Fabio Petroni, Aleksandra Piktus, Angela Fan, Patrick Lewis, Majid Yazdani,
  Nicola De~Cao, James Thorne, Yacine Jernite, Vladimir Karpukhin, Jean
  Maillard, Vassilis Plachouras, Tim Rockt{\"a}schel, and Sebastian Riedel.
  2021.
\newblock \href {https://doi.org/10.18653/v1/2021.naacl-main.200} {{KILT}: a
  benchmark for knowledge intensive language tasks}.
\newblock In \emph{Proceedings of the 2021 Conference of the North American
  Chapter of the Association for Computational Linguistics: Human Language
  Technologies}, pages 2523--2544, Online. Association for Computational
  Linguistics.

\bibitem[{Vaswani et~al.(2017)Vaswani, Shazeer, Parmar, Uszkoreit, Jones,
  Gomez, Kaiser, and Polosukhin}]{DBLP:journals/corr/VaswaniSPUJGKP17}
Ashish Vaswani, Noam Shazeer, Niki Parmar, Jakob Uszkoreit, Llion Jones,
  Aidan~N. Gomez, Lukasz Kaiser, and Illia Polosukhin. 2017.
\newblock \href {http://arxiv.org/abs/1706.03762} {Attention is all you need}.
\newblock \emph{CoRR}, abs/1706.03762.

\bibitem[{Yang et~al.(2019)Yang, Lin, Wang, Lin, and Tsai}]{Yang2019QueryAA}
Jheng-Hong Yang, Sheng-Chieh Lin, Chuan-Ju Wang, Jimmy Lin, and Ming-Feng Tsai.
  2019.
\newblock Query and answer expansion from conversation history.
\newblock In \emph{TREC}.

\end{thebibliography}
\bibliographystyle{acl_natbib}

\end{document}